\documentclass[a4paper,twoside]{article}
\newcommand{\affil}[1]{$^{\rm #1}$}
\usepackage[authoryear]{natbib}
\bibpunct{(}{)}{;}{a}{}{,}
\usepackage{graphicx}
\date{} 
\newcommand{\kms}{\mbox{km\,s$^{-1}$}}
\def\kms{km${\rm s}^{-1}$}
\def\arcdeg{\hbox{$^\circ$}}
\def\arcmin{\hbox{$^\prime$}}
\def\arcsec{\hbox{$^{\prime\prime}$}}
\def\ergcms{erg\,cm$^{-2}$s$^{-1}$}
\def\ha{H$\alpha$}

\def\NII{[N\,\textsc{ii}]}
\def\OIII{[O\,\textsc{iii}]}
\def\HII{H\,\textsc{ii}}

\def\p0{\phantom{0}}
\def\lessim{\raise-.5ex\hbox{$\buildrel<\over\sim$}}
\def\grtsim{\raise-.5ex\hbox{$\buildrel>\over\sim$}}

\title{\large\bf\flushleft K~1-6: an asymmetric planetary nebula with a  binary central star}
\author{\parbox{\textwidth}{\flushleft
\vspace{-0.5cm}
{\it David J. Frew\affil{A,I}, Jeff Stanger\affil{B}, Michael Fitzgerald\affil{A}, Quentin Parker\affil{A,C}, Lena Danaia\affil{D}, David McKinnon\affil{D}, Mart\'in A. Guerrero\affil{E,F}, John Hedberg\affil{G}, Robert Hollow\affil{H},   Yvonne An\affil{B}, Shu Han Bor\affil{B}, Isabel Colman\affil{B}, Claire Graham-White\affil{B}, Qing Wen Li\affil{B}, Juliette Mai\affil{B}, Katerina Papadakis\affil{B}, Julia Picone-Murray\affil{B}, Melanie Vo Hoang\affil{B}, and Vivian Yean\affil{B}}\\
\vspace{0.4cm}
{\small \affil{A}\,Department of Physics and Astronomy, Macquarie University, North Ryde, NSW 2109, Australia}\\
{\small \affil{B}\,Sydney Girls High School, Anzac Parade, Surry Hills, NSW 2010, Australia}\\
{\small \affil{C}\,Australian Astronomical Observatory, PO Box 296, Epping, NSW 1710, Australia}\\
{\small \affil{D}\,School of Teacher Education, Charles Sturt University, Bathurst, NSW 2795, Australia}\\
{\small \affil{E}\,Instituto de Astrof\'{\i}sica de Andaluc\'{\i}a, CSIC, c$/$Camino Bajo de Hu\'etor 50, E-18008 Granada, Spain}\\
{\small \affil{F}\,Visiting Scholar, Macquarie University, North Ryde, NSW 2109 Australia}\\
{\small \affil{G}\,Department of Education, Macquarie University, North Ryde, NSW 2109 Australia}\\
{\small \affil{H}\,CSIRO Astronomy and Space Science, PO Box 76, Epping, NSW 1710 Australia}\\
{\small \affil{I}\,Email: david.frew@mq.edu.au}}}

\begin{document}
\twocolumn[
\begin{changemargin}{.8cm}{.5cm}
\begin{minipage}{.9\textwidth}
\vspace{-1cm}
\maketitle
%
%
\small{\bf Abstract:}

We present new imaging data and archival multiwavelength observations of the little studied emission nebula K~1-6 and its central star.  Narrow-band images in \ha\ (+ \NII) and \OIII\ taken with the Faulkes Telescope North reveal a stratified, asymmetric, elliptical nebula surrounding a central star which has the colours of a late~G or early K-type subgiant or giant.  GALEX ultraviolet images reveal a very hot subdwarf or white dwarf coincident in position with this star.   The cooler, optically dominant star is strongly variable with a period of 21.312 $\pm$ 0.008 days, and is possibly a high amplitude member of the RS~CVn class, although an FK~Com classification is also possible.  Archival ROSAT data provide good evidence that the cool star has an active corona.  We conclude that K~1-6 is most likely an old bona fide planetary nebula at a distance of $\sim$1.0\,kpc, interacting with the interstellar medium, and containing a binary or ternary central star.  The observations and data analyses reported in this paper were conducted in conjunction with Year~11 high school students as part of an Australian Research Council Linkage Grant science education project, denoted Space To Grow,  conducted jointly by professional astronomers, educational researchers, teachers, and high-school students.  



\medskip
\medskip
\end{minipage}
\end{changemargin}
]
\small

\section{Introduction}
\label{sec:intro}


Planetary nebulae (PN) are an important, albeit brief ($\leq$10$^5$ yr) phase in the evolution of low- to intermediate-mass stars ($\sim$1 to 8 M$_{\odot}$).  They are formed at the end of the asymptotic giant branch (AGB) phase, as the red giant ejects its outer envelope in a slow but massive `superwind'.  After the envelope ejection, the remnant core of the star increases in temperature before nuclear burning ceases and the star fades to become a white dwarf.   The hot surface temperature of the central star causes the previously ejected material to be ionized, which becomes visible as a PN, its shape influenced by the interaction between the old red giant envelope and a tenuous, fast wind from the hot central star (Kwok, Purton \& Fitzgerald 1978).

Recent work has shown a remarkable diversity in PN morphologies and central star properties (Frew \& Parker 2010).  Since evidence is increasing that the `PN phenomenon' is heterogeneous, and PNe are likely to be formed from multiple evolutionary scenarios, it is important to ascertain the nature of any poorly studied emission nebula that has been ambiguously classified in the literature.

The object known as K~1-6 is a case in point.  It was discovered by Kohoutek (1962), who noted a faint nebulosity visible on Palomar Observatory Sky Survey (POSS-I) red and blue prints.  He classed it as a probable PN, even though he had no spectroscopic data, nor did he find a candidate ionizing star.  Due to its low surface brightness, it has been little studied over the half century since its discovery.  Perek \& Kohoutek (1967) included it in their {\it Catalogue of Galactic Planetary Nebulae} under the designation PK~107+21.1.  Cahn \& Kaler (1971), assuming it was a PN, estimated a red flux from the POSS.  Kwitter, Jacoby \& Lydon (1988) took CCD images centred on the nebula, but also failed to find an ionizing star.  Consequently, its status as a bona fide PN was in doubt.  Acker et al. (1992) classed it only as a possible PN, with the comment that it was a faint, highly reddened object showing weak \ha\ emission, based on unpublished spectra.  On the other hand, Kohoutek (2001) included it as a bona fide PN, without elaborating on any reasons for its classification.	 

K 1-6 is visible on both blue and red POSS-II plates, but reflection nebulae are also known to occur at high Galactic latitudes (e.g. Goerigk et al. 1983).  Since it is a high latitude object at a very northerly declination (+74\arcdeg) we cannot use the IPHAS (Drew et al. 2005) or SHS (Parker et al. 2005) surveys to investigate its morphology on deep \ha\ images.  However, it is visible on a low-resolution (96\arcsec), continuum-subtracted \ha\ image downloaded from the Virginia Tech Spectral line Survey\footnote{http://www.phys.vt.edu/$\sim$halpha/} (VTSS; Dennison, Simonetti \& Topasna 1998), from which we can conclude it is a genuine emission nebula.  Since the latitude is quite high, an informed guess would class it as a PN rather than a conventional \HII\ region.  However, deep, higher resolution, narrowband images are needed to better understand its morphology, as well as evidence that a low-luminosity ionizing star is present in the vicinity before it can be definitively classified as a PN.


In this paper, we describe our efforts to ascertain the nature of K~1-6, selected as an object worthy of study as part of our \emph{Space to Grow} project.  This is a science education project conducted jointly by professional astronomers, educational researchers, teachers, and high-school students supported by an Australian Research Council (ARC) Linkage Grant.   In Section~\ref{sec:ARC} we give some background on this science education project, in Section~\ref{sec:observations} we describe new narrowband imaging data acquired with the Faulkes Telescope North, as well as archival multiwavelength data obtained from the Virtual Observatory (VO).  We discuss our interpretations on the nature of K~1-6 in Section~\ref{sec:analysis}, offer suggestions for future work in Section~\ref{sec:future}, before summarising our conclusions in Section~\ref{sec:conclusions}.

\section{The Faulkes Telescope ARC-linkage Project}\label{sec:ARC}

The \emph{Space to Grow} Faulkes Telescope Project is a three year ARC Linkage Grant that uses astronomy as the vehicle to engage students in real science during their final years of secondary school (Years 10--12).  Launched in 2009, this major education project is administered by Macquarie University and follows the success of our Australian School Innovation in Science, Technology and Mathematics (ASISTM) pilot project\footnote{http://www.astronomy.mq.edu.au/asistm/} to help address the serious problem of poor student engagement and retention in science in Australian high schools.  The industry partners in this ARC  linkage project are Charles Sturt University, the Catholic Schools Offices of Parramatta and Bathurst, the NSW Department of Education (Western Region) and the Las Cumbres Observatory Global Telescope network (LCOGT).\footnote{http://lcogt.net/}  

The project is funded until 2012, and we will potentially have up to 40 schools, 200 teachers, and approximately 5000 students from across different educational sectors in NSW involved in the project.   Students and their teachers have the opportunity to work with professional astronomers on existing projects and create their own research proposal and submit requests to obtain data from the two, 2-metre Faulkes Telescopes located in Coonabarabran, NSW and Maui, Hawaii.  These are the world's largest optical telescopes regularly available for science education projects in schools.

A suite of detailed learning materials (accessible at the Space To Grow Website\footnote{http://www.astronomy.mq.edu.au/space2grow/}) is being developed to guide students' learning and help them develop the necessary skills to both apply for telescope time and analyse the scientific data obtained from such observations. In addition, ongoing teacher professional development is embedded within the project to support teachers in competently covering the astronomy content of the NSW science curriculum and to give them the confidence to use technology with their students in science lessons.
 
A similar project exists in the context of radio astronomy, the CSIRO's PULSE$@$Parkes Project\footnote{http://outreach.atnf.csiro.au/education/pulseatparkes/} where students have the opportunity to control the 64-m Parkes radio telescope to observe galactic pulsars, analyse their own data on these objects, and contribute to professional astronomers' science projects (Hollow et al. 2008; Hobbs et al. 2009).  The overall aim of these projects is to engage students in real science and to support teachers in dealing with the processes and approaches that motivate science students. Both the Space to Grow and the PULSE$@$Parkes projects have embedded educational research procedures to investigate the impact of these science experiences on both students and teachers.

The observations and data analyses reported in this paper were conducted in conjunction with Year~11 high school students with the goal of involving them in real science.  One of us (J.S.) is a science teacher who was originally involved in the ASISTM project that provided students with access to the Faulkes Telescopes.  We emphasise that the students were embedded in the ``journey of scientific discovery'' throughout the study of this interesting nebula, with in-depth discussion sessions conducted as part of school visits by two team members (D.J.F. and M.F.).  More specifically, the students generated the colour composite image of the PN using various software packages, and, in conjunction with their teacher (JS), used online tools to retrieve the various archival data sets utilised in the paper, before generating the light curves for the central star, estimating the period and amplitude, and investigating the variability type based on the available data.  We emphasise that all software packages used in the analysis are freely available to high school teachers and their students.

\section{Observations}\label{sec:observations}

\subsection{Faulkes Telescope Images}


Narrowband images of the nebula were obtained with the 2-metre f/10 Faulkes Telescope North (FTN) at Haleakala Observatory, Maui.  The Merope camera was used which contains an E2V CCD with 2048\,$\times$\,2048 pixels.  The plate scale of 0.14 arcsec\,pixel$^{-1}$ leads to a field-of-view of 4.7\,arcmin on a side.   Two preliminary 600\,sec exposures in each of two narrowband filters centred on the \ha\ $\lambda$6563\AA\ and \OIII\ $\lambda$5007\AA\ lines were acquired on the night of 30 November 2009, with the goal of better understanding the nebular morphology. Note that the \ha\ filter ($\lambda_{\rm eff}$ = 6560\AA, FWHM = 92\AA) transmits the two red \NII\ lines in the wings of the filter. Shorter exposures were also taken in the standard Johnson $B$ and $V$ filters.  Prior to downloading, the raw images were pipeline-processed at LCOGT which included standard flat-fielding and bias subtraction routines (Lewis et al. 2010).  The resultant images were then retrieved as 2$\times$2 binned FITS format images from the LCOGT website. 

The observing conditions on the night were not optimal however, as the frames were taken through a large airmass with poor seeing, and due to the long exposure times at large zenith angle, the images suffered from modest smearing due to tracking error and/or wind.  Consequently, we obtained an additional 27 five minute exposures in  \OIII\ on 17/18 April 2010 and 15 five minute \ha\ exposures on 8 May 2010.  Images were rejected if they had visible cosmetic defects or flat-fielding problems.  We finally used 18 individual \OIII\ frames and 10 \ha\ frames for generating deeper images with higher S/N.   The individual images were aligned in the Astrometrica\footnote{http://www.astrometrica.at/} software package using an object detection and matching algorithm, and then median stacked to eliminate bad pixels and cosmic ray hits.  The resulting final images are presented and discussed in \S\ref{sec:analysis}.

\subsection{Archival Data}

To supplement our imaging data, we searched for multi-wavelength archival data of the nebula and central star using Virtual Observatory\footnote{http://www.ivoa.net/} tools, accessible to professional and amateur astronomers, educators, and students.  Most of the data were retrieved and visualised primarily via the Aladin Sky Atlas from the Centre de Donn\'ees astronomiques de Strasbourg (CDS), and the SkyView Virtual Observatory\footnote{http://skyview.gsfc.nasa.gov/}.    Oostra (2006) has discussed the utility of using online databases for astronomical research in the classroom.  

\begin{figure*}
\begin{center}
\includegraphics[width=5.2cm]{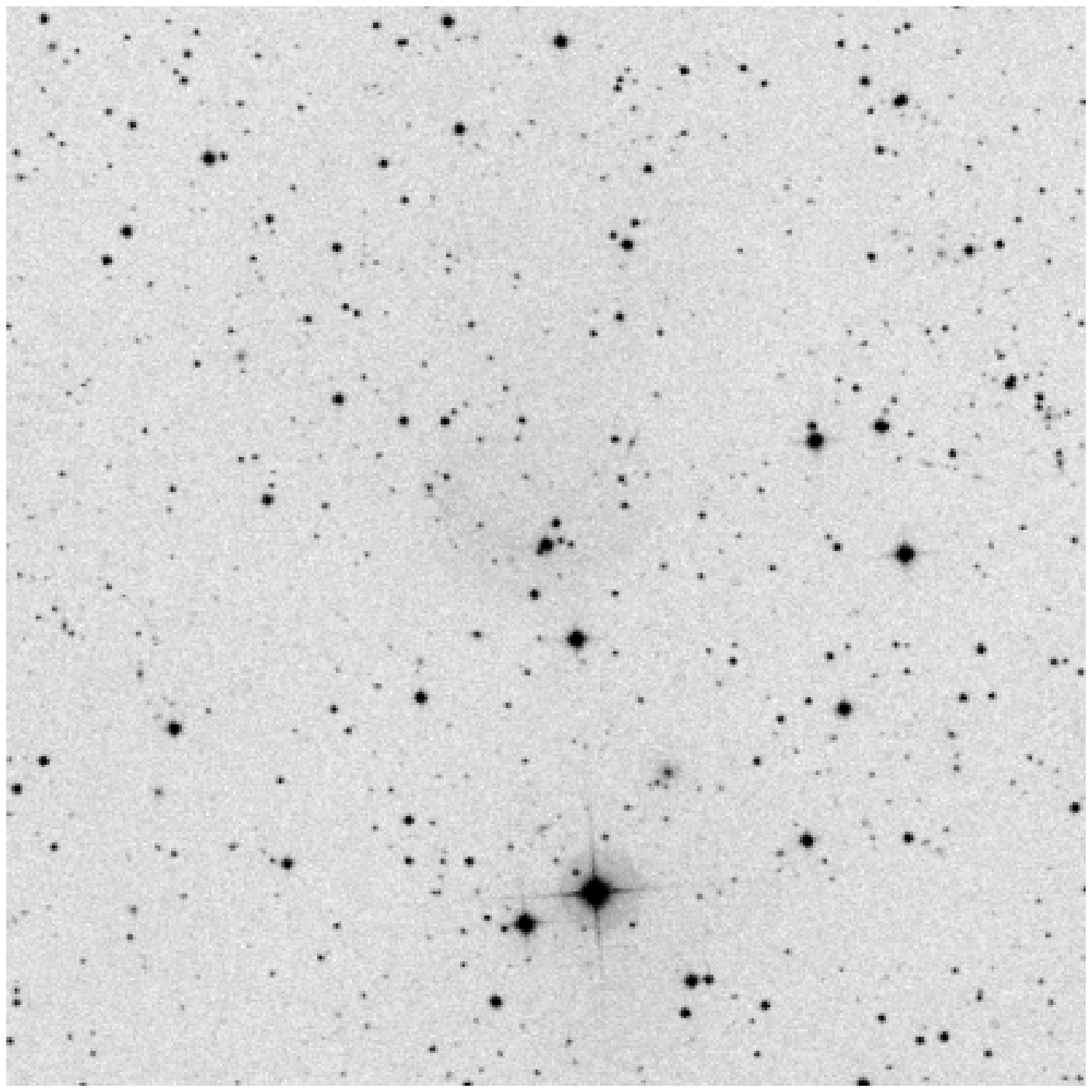}
\includegraphics[width=5.2cm]{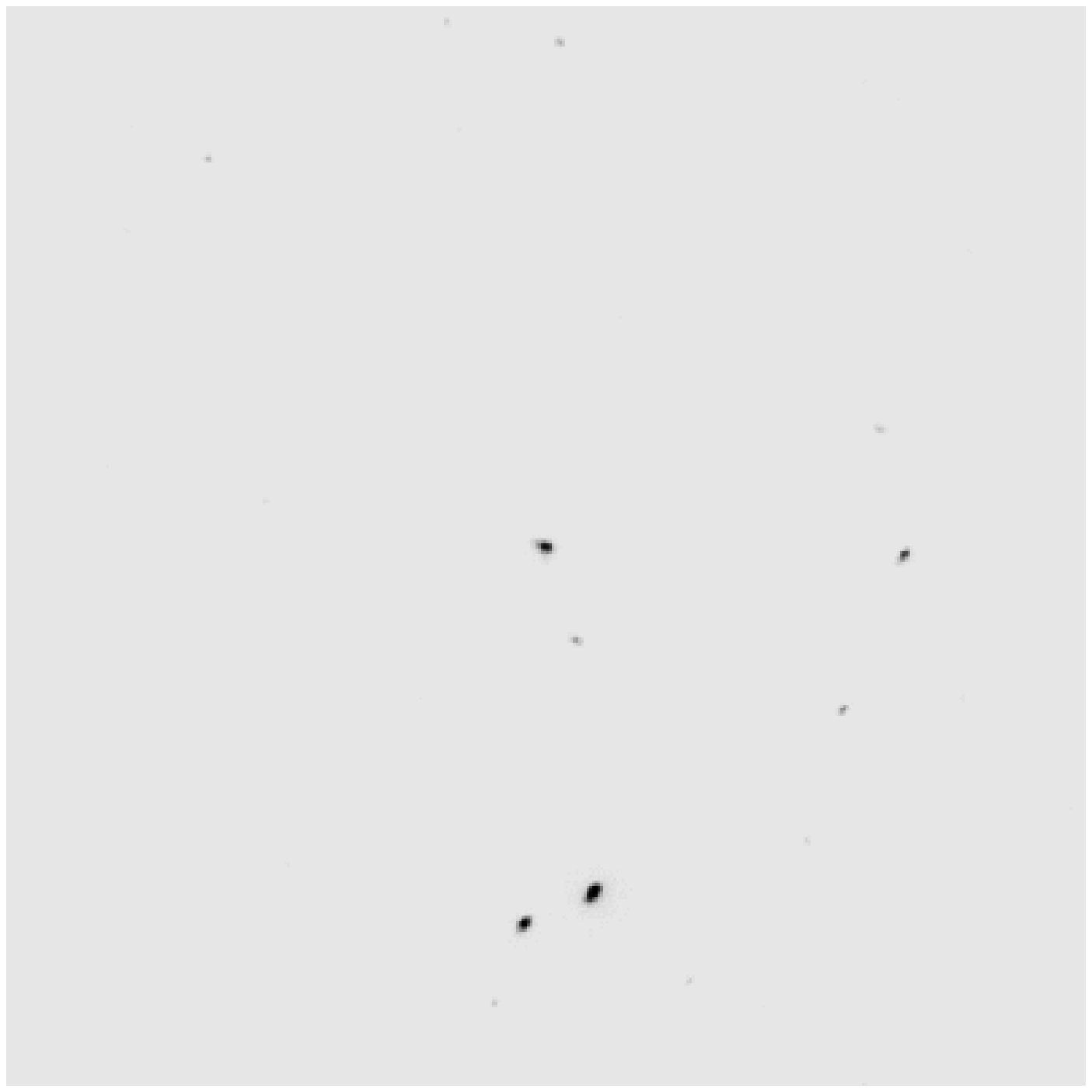}
\includegraphics[width=5.2cm]{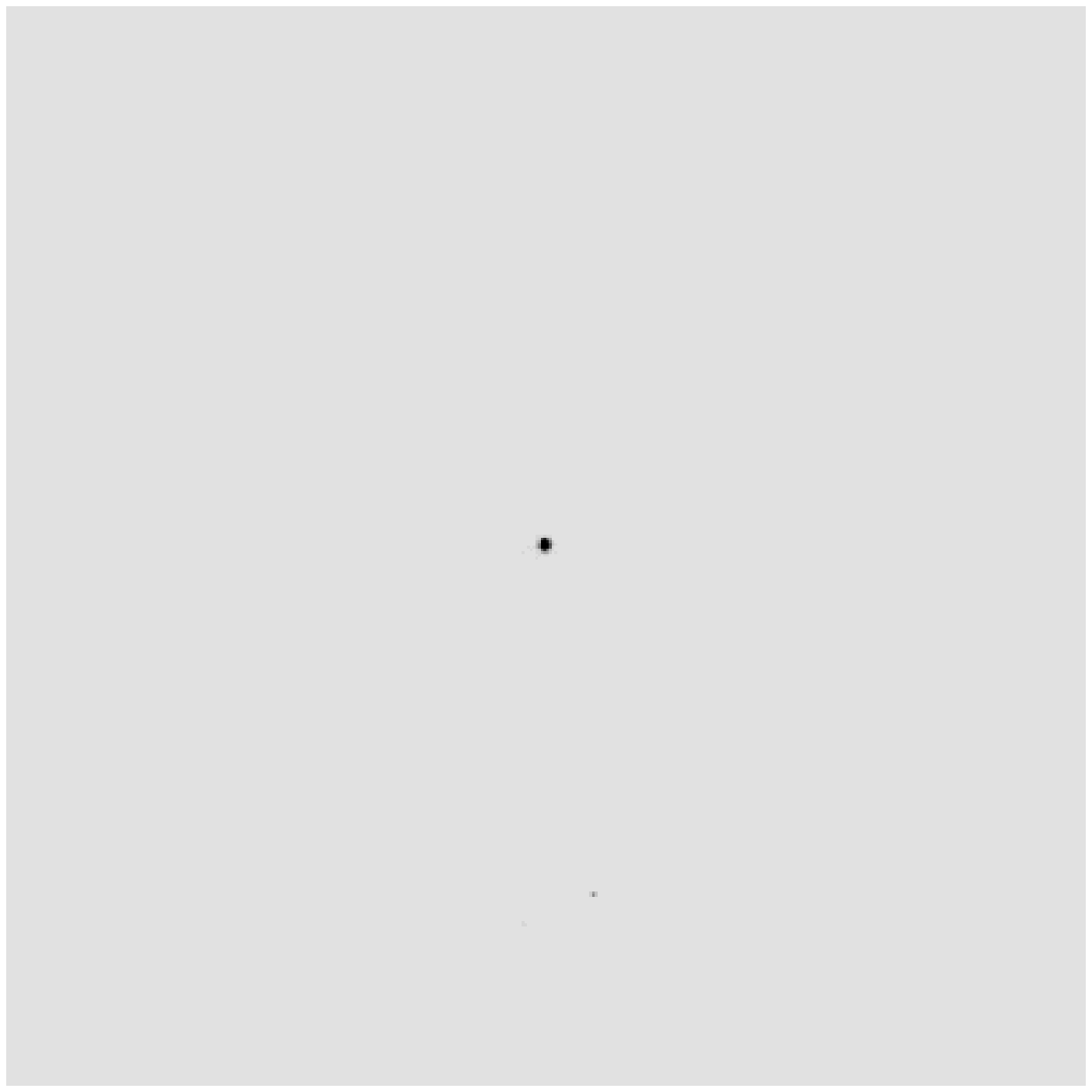}
\caption{Left:  Red DSS2 image of K~1-6, showing the faint nebula at centre.  Middle: Corresponding GALEX NUV image of the same field. Right:  Corresponding GALEX FUV image, showing the prominent, very hot ionizing source.  Each image is 15\arcmin\ on a side, with north at top.}
\label{fig:K1-6_GALEX}
\end{center}
\end{figure*}

We used archival images from the Galaxy Evolution Explorer (GALEX) satellite to look for a hot source in the vicinity, as our central star candidate is too cool to ionize the nebula.   GALEX is a space-based 50\,cm telescope currently undertaking a nearly all-sky imaging survey using two ultraviolet (UV) filters: an FUV band (covering 1350--1750 \AA) and an NUV band (1750--2750 \AA).   Further details of the survey are given in Martin et al. (2005) and Morrissey et al. (2007).   We visualised images downloaded via the GalexView 1.4 web-based utility\footnote{http://galex.stsci.edu/galexView/}.   A prominent, very hot star (FUV flux $>$ NUV flux) was noted on Field~AIS\_46, designated GALEX J200414.5+742535 (see Figure~\ref{fig:K1-6_GALEX}).  The coordinates of this source from GALEX are within 2\arcsec\ of our candidate central star.  The positions are formally identical as the GALEX angular resolution is 4.5--6\arcsec\ (Martin et al. 2005).   We emphasise that no other conspicuous FUV source is visible inside a 15\arcmin\ $\times$ 15\arcmin\ region centred on K~1-6 (see Figure~\ref{fig:K1-6_GALEX}), so this star is clearly the ionizing source for the emission nebula.  Our central star candidate is the optically brightest star visible in the interior of the emission nebula seen on our FTN images, and as we show below, has the colours of a G- or K-type star.  We adopt its position from the Naval Observatory Merged Astrometric Dataset (NOMAD)\footnote{http://www.nofs.navy.mil/nomad} accessed through the VizieR\footnote{http://vizier.u-strasbg.fr/cgi-bin/VizieR} service at the CDS.  We also adopt these coordinates as the nebular position (see Table~\ref{table:basic}).

Once the central star was identified, additional data was retrieved using VizieR, using a 5\arcsec\ search radius around the nominal position from NOMAD.  Additional photometric data was obtained from the Tycho-2 catalogue (H$\o$g et al. 2000) and the Two Micron All Sky Survey (2MASS; Cutri et al. 2003; Skrutskie et al. 2006), supplemented by lower-precision magnitudes from the various scanned Schmidt plate surveys.  Time-series photometric data were obtained from the Northern Sky Variability Survey website\footnote{http://skydot.lanl.gov/nsvs/nsvs.php} (NSVS; Wo\'zniak et al. 2004) and The Amateur Sky Survey (TASS; Droege et al. 2006) webpage\footnote{http://www.tass-survey.org/}.

We also noted an X-ray detection in the ROSAT All-Sky Bright Source Catalogue (Voges et al. 1999) with the designation 1RXS\,J200413.0+742533 (Figure~\ref{fig:ROSAT}).  This object is coincident in position with both the cool star seen on the DSS and the GALEX source to within the nominal errors.  The source was observed by ROSAT PSPC for 2350~sec (Seq.\,ID RS930419N00) at a count rate of 0.035 $\pm$ 0.006 counts~s$^{-1}$, thus resulting in 83 $\pm$ 15 backgound-subtracted counts.  Further implications of the associated X-ray source are discussed in Section~\ref{sec:x-ray}.

Finally, we note the absence of a nebular radio detection in the survey of Zijlstra, Pottasch \& Bignell (1989), who gave an upper limit of 4\,mJy at 1.5\,GHz.  We visualised a 20\arcmin\ NRAO VLA Sky Survey (NVSS) image (Condon et al. 1998) using SkyView and confirm this result.  An unrelated background source is located 2.3\arcmin\ north-west of the nebular position.

\section{Analysis}\label{sec:analysis}

\subsection{Nebular Morphology}\label{sec:morphology}

\begin{figure*}
\begin{center}
\includegraphics[width=5.1cm]{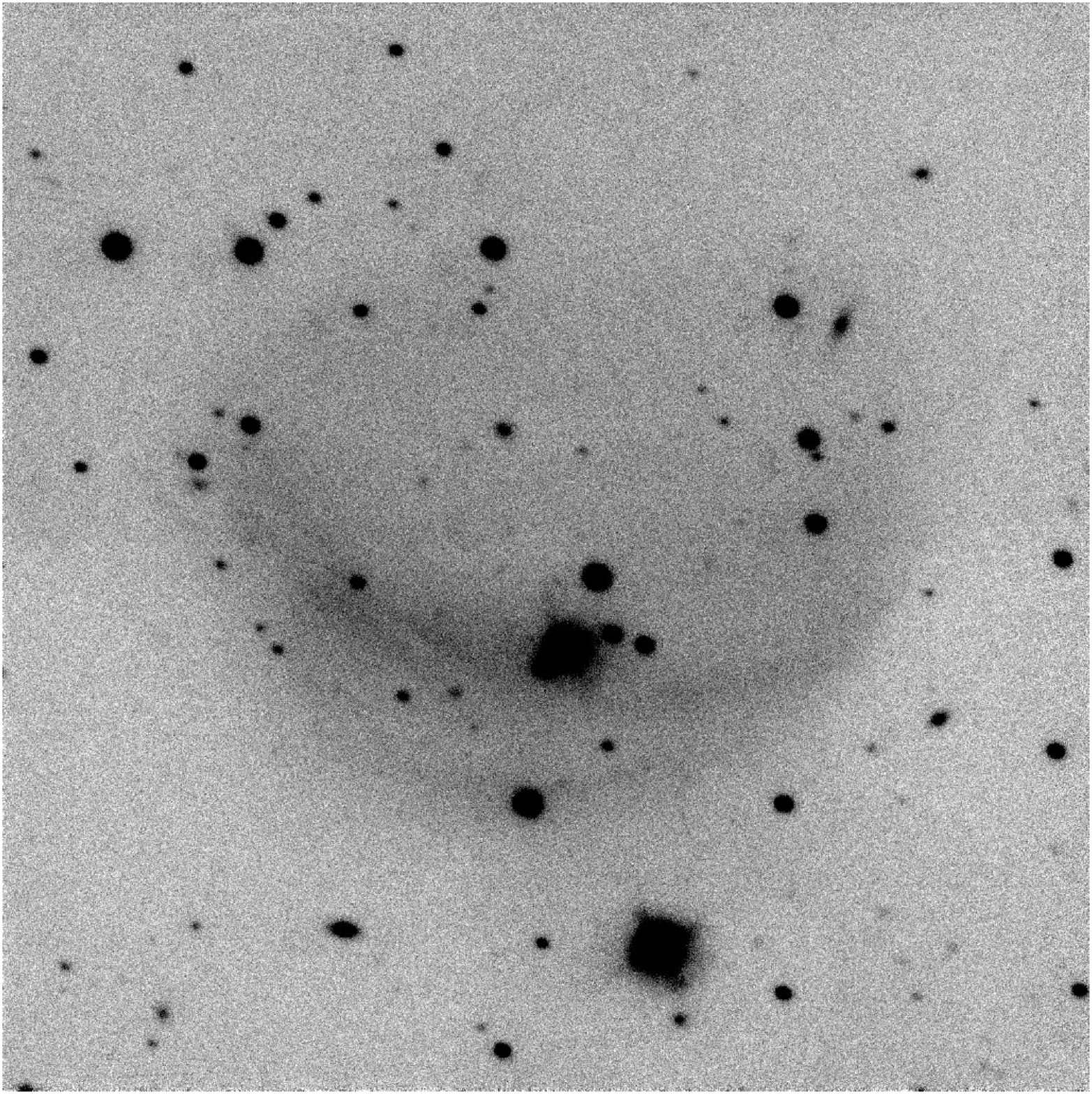}
\includegraphics[width=5.1cm]{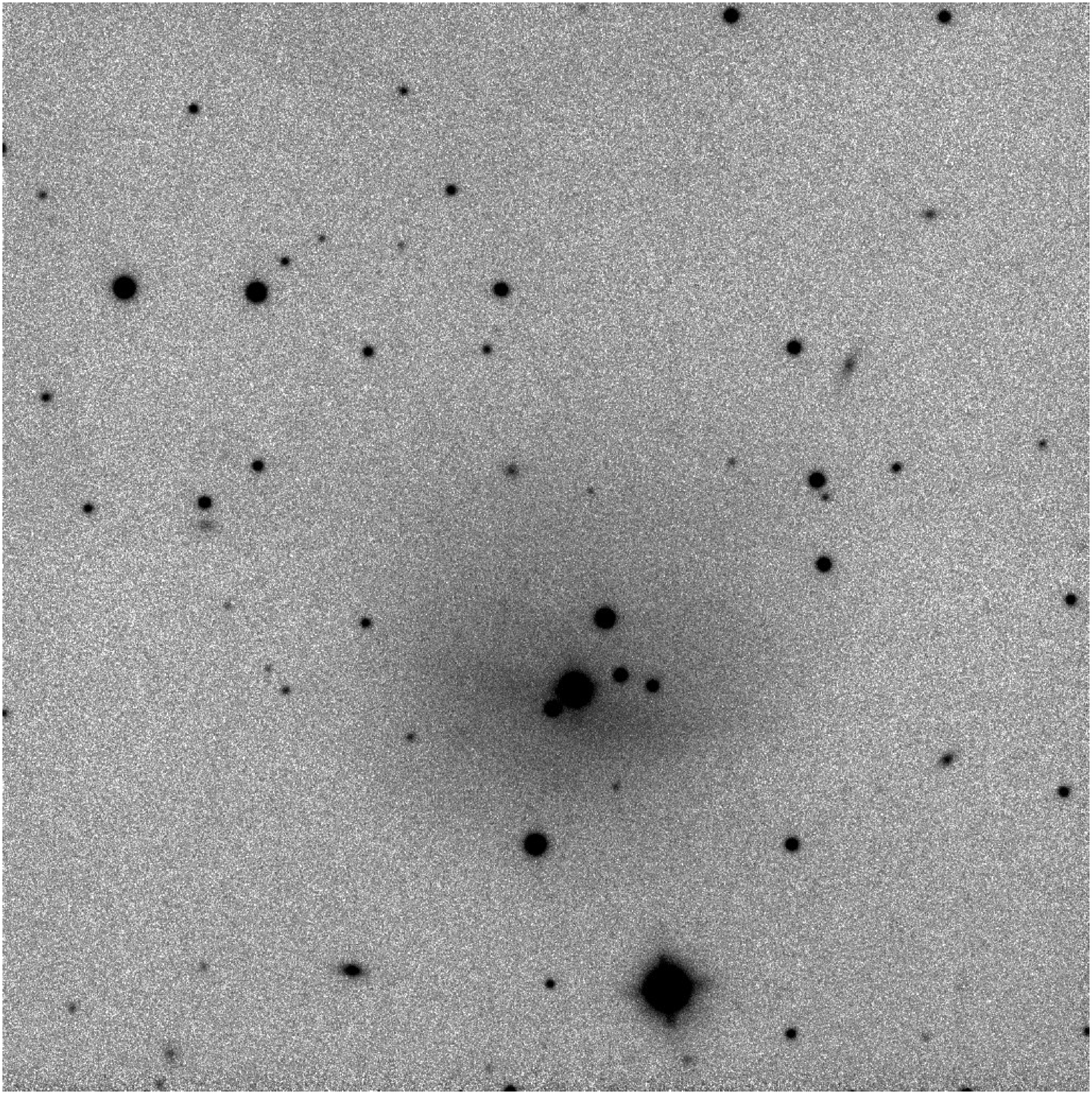}
\includegraphics[width=5.45cm]{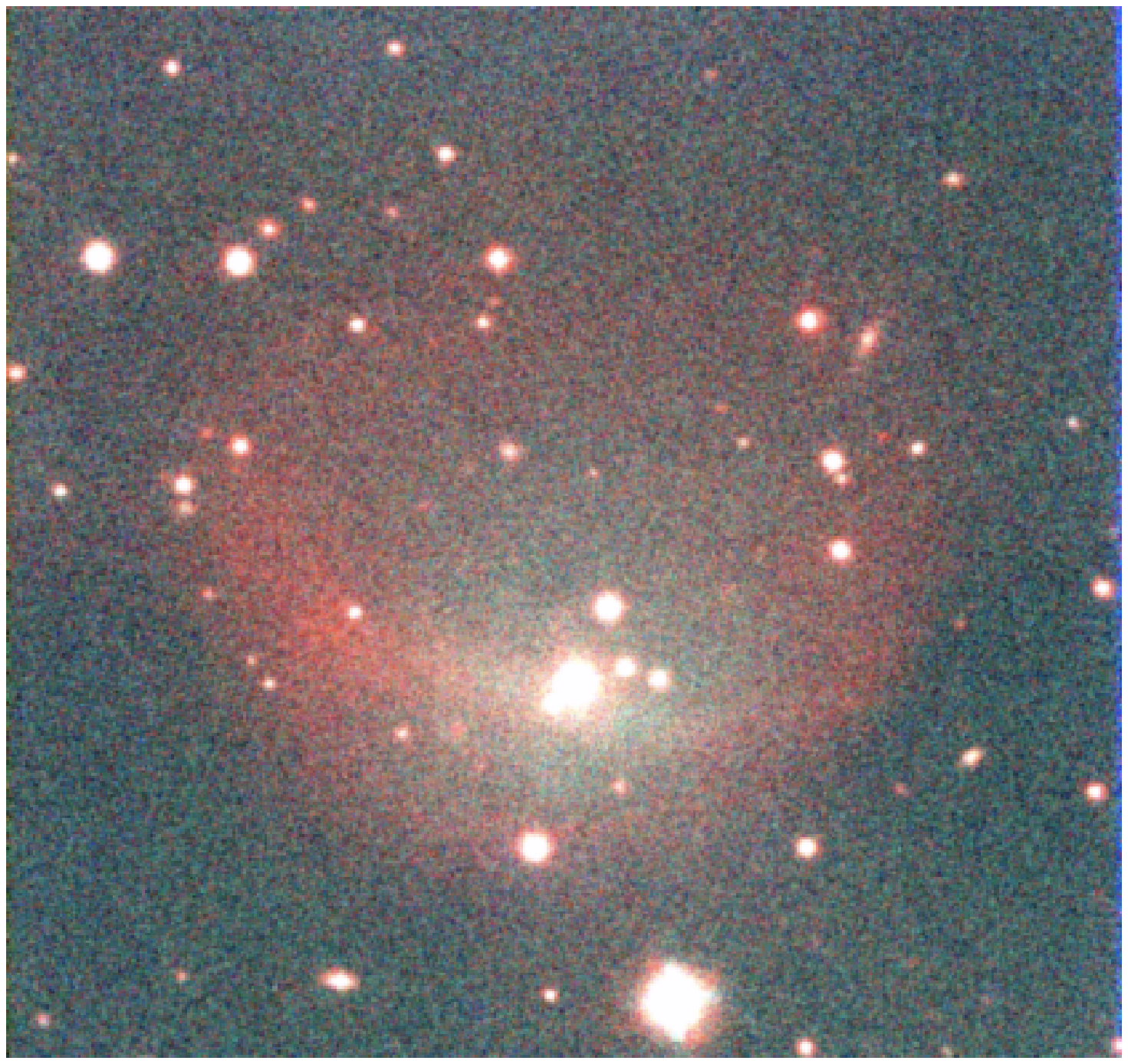}
\caption{Left: FTN \ha\ image of the one-sided, evolved PN K~1-6.  Middle: FTN \OIII\ image of K~1-6. Right:  Colour image combined from H$\alpha$ (red), \OIII\ (blue) and combined H$\alpha$ + \OIII\ (green).  The central star is marked.  The closest of several unrelated field stars is 7.2\arcsec\ SE of the central star, and partly blended with it on these images.  Each image is 4.7\arcmin\ across, with north-east at top left.}
\label{fig:K1-6_FTN}
\end{center}
\end{figure*}

We present our narrowband and colour-composite images of K~1-6 in Figure~\ref{fig:K1-6_FTN}.  The \ha\ + \NII\ image shows the nebula to have an asymmetric appearance, with a brighter rim on the south side, while the \OIII\ image shows the O$^{2+}$ emitting zone to be roughly circular and restricted to a zone around the best candidate for the central star, i.e. the nebula shows ionization stratification.  Similar stratification is commonly seen in evolved PNe with faint central stars on the white dwarf cooling track.    The dimensions of the nebula were measured from the combined FTN \ha\ image as 198 $\times$ 160\arcsec.   The available observational data on the nebula are summarised in Table~\ref{table:basic}.

The position of our candidate central star is offset from the nebular centroid, closer to the brighter rim of the nebula as seen in \ha\ + \NII\ light (see Figure~\ref{fig:K1-6_FTN}).  We explain this  arcuate morphology and the location of the ionizing source as a product of an nebular interaction with the interstellar medium (ISM) as discussed by Wareing et al. (2007),  Wareing (2010) and Sabin et al. (2010).   We discuss the PN/ISM interaction in more detail in Section~\ref{sec:interaction}.

On our short-exposure FTN $B$ images, the cool star appears perfectly round, showing that the hot star must have a separation of $\leq$1.2\arcsec\ (if it has V$\simeq$15.8; see below). There are also five unrelated field stars within 30\arcsec\ of the central binary; the closest has $V$\,$\sim$\,16 and is 7.2\arcsec\ south-east (see Figure~\ref{fig:K1-6_FTN}).  Since the positions of the cool and hot components of the central star are identical within the errors, we can state that the optically cool star has a very strong UV-excess and is potentially part of an unresolved binary system.  In addition, we find the cool component to have a significant proper motion.  The values from the USNO CCD Astrograph Catalog (UCAC3; Zacharias et al. 2010), the Tycho-2 catalogue and the USNO-B catalogue (Monet et al. 2003) are highly consistent, especially in the $\delta$ component.  The values are much larger than the formal errors, and we take the detection to be real and not spurious.  We adopt the UCAC3 value of $\mu$ = 29.1~mas\,yr$^{-1}$ in position angle 179\arcdeg.

This motion is practically due south, and is perfectly consistent with the morphology of the bow shock on the southern edge of the nebula seen on our FTN \ha\ images.  This is a key piece of evidence to support our hypothesis that \emph{the cool star is a physical companion to the true ionizing star}.  While the GALEX astrometry is not precise enough to rule out a chance superposition, we consider it improbable for the cool star to have a proper motion vector in the correct orientation, without it being associated with the PN.  The asymmetry in the PN shell also leads to the conclusion that the nebula was ejected from the central binary system, and is not ionized ambient ISM (see Frew \& Parker 2010).  We assume hereafter that this is an unresolved binary system, where the hot component is ionizing the surrounding nebula. Recall that we place an upper limit of 1.2\arcsec\ on the separation of the components from our FTN $B$ short exposure images.

\subsection{K~1-6, a bona fide PN}\label{sec:nature}

Is K 1-6 is a highly evolved PN?  We note that PN-like nebulae with parabolic bow-shock morphologies are known (see Frew \& Parker 2010, for a review).  These include pulsar-wind nebulae and bowshocks associated with nova-like cataclysmic variables (CVs), e.g. EGB~4 (Ellis, Grayson \& Bond 1984; Krautter, Klaas \& Radons 1987; Hollis et al. 1992) and Fr~2-11 (Frew, Madsen \& Parker 2006; Frew et al. 2010b).  The peculiar nebula Abell~35 (Jacoby 1981) has a very similar morphology to Fr~2-11 and is now thought to be a bowshock nebula inside a photoionized ambient \HII\ region (Frew 2008).  In both these latter cases, the bowshock is only visible in the light of \OIII\ and is interior to the external edge of the ionized region.  The observed morphology of K~1-6 is different and similar to other one-sided PN; recall Figure~\ref{fig:K1-6_FTN} which shows that K~1-6 has an exterior bowshock which implies the ionized nebula is moving with respect to the ISM (cf. the emission nebula ionized by PHL~932; Frew et al. 2010a).

\begin{table}
{\footnotesize
\begin{center}
\caption{Basic properties of the nebula, K~1-6.}
\medskip
\label{table:basic}
\begin{tabular}{ll}
\hline
$\alpha$ (J2000)                   & 20 04 14.25\\				
$\delta$ (J2000)	          & +74 25 36.4 \\				
$l$	                       & 107.036    \\				
$b$	                       & +21.384    \\				
Dimensions & 198\arcsec $\times$ 160\arcsec \\
log F(H$\alpha$)  (cgs)              & $-11.32$ $\pm$ 0.07     \\				
$E(B-V)$                           &  0.22       \\
$c_{\rm H\beta}$                    &  0.32       \\
Distance                            &  $\sim$1.0 kpc  \\   
$z$-height                          &  $\sim$365 pc  \\ 
\hline
\end{tabular}
\end{center}
}
\end{table}

We have calculated the ionized mass of the nebula using equation~7 from Hua \& Kwok (1999), which uses the measured diameter and integrated \ha\ flux, and an estimate of the distance, taken from Section~\ref{sec:distance}:  

\begin{equation} 
M_{\rm ion} = 0.032\, (\epsilon/0.6)^{0.5} \theta^{1.5} D^{2.5} F_{0}({\rm H}\alpha)^{0.5} \, M_{\odot}
\end{equation}

where $\theta$ is in arcmin, $D$ is in kpc, and $F_{0}({\rm H}\alpha)=F({\rm H}\alpha)\times 10^{0.69\times c_{{\rm H}\beta}}$ is in units of 10$^{-12}$ erg\,cm$^{-2}$\,s$^{-1}$.  We assumed a volume filling factor, $\epsilon$ = 0.6 following Hua \& Kwok (1999), and determine the integrated \ha\ flux in the following manner. No PN flux standards were observed, so we perform aperture photometry from the calibrated VTSS \ha\ image (Dennison et al. 1998), following Frew, Parker \& Russeil (2006).  We measure  a \ha\ flux of 4.8 $\pm$ 0.8 $\times$ 10$^{-12}$ \ergcms\ through a 7\arcmin\ aperture.  No correction for \NII\ emission is needed as these lines are not passed by the narrowband VTSS filter, but we have corrected for the known zero-point error of the VTSS images (Gaustad et al. 2001; Finkbeiner 2003).  The log flux value is presented in Table~\ref{table:basic}.

At a distance of 1.0~kpc, the ionized mass of K~1-6 is 0.15~$M_{\odot}$, consistent with other PNe, which typically have ionized masses ranging from 5$\times$10$^{-3}$ to 3 $M_{\odot}$ (Frew \& Parker 2010).  It is far too massive to be an evolved nova shell, which usually has a mass of the order of 10$^{-4}$~$M_{\odot}$ (Cohen \& Rosenthal 1983).  In addition, the observed morphology is quite unlike that of a nova shell.

Based on our new images, K~1-6 shows no evidence of any bipolarity or axisymmetric, low-ionization structures (LIS; Gon\c{c}alves, Corradi \& Mampaso 2001) which may be indicative of a {\it close binary} nucleus (see Miszalski et al. 2009, 2010a).   The influence of binarity on PN formation and shaping is an important topic in the field, and is currently being debated (see De Marco 2009 for a recent review).  Hence, any new binary central stars with periods in the weeks to years range are very important.  Since we do not yet have any spectroscopic data on the central star, we have no information on the separation and orbital period of the binary components, but it is tempting to identify the 21.3 day photometric period (see Section~\ref{sec:variability}) as the orbital period of the system.  However, such a system would be unusual in a non-bipolar PN.

In summary, our FTN narrowband imaging data reveal a faint elliptical nebula that shows an asymmetric morphology indicative of an interaction with the ISM.  The available data suggest it is best interpreted as an evolved, low-surface-brightness PN based on its bow-shocked morphology, ionization structure, and ionized mass. Following the nomenclature introduced by Acker et al. (1992), the nebula is designated as PN~G107.0+21.3.

\subsection{Stellar photometry}\label{sec:photometry}


We summarise the available literature photometry of the binary central star in Table~\ref{table:K1-6_CS}.   In order to correct the observed colours of the cool star for reddening, a colour excess of E$(B-V)$ = 0.22 mag is adopted from the dust maps of Schlegel, Finkbeiner \& Davis (1998) on this sightline, though this is technically an upper limit.  The visual absorption is then $A_v$ = 0.68 mag using the $R$ = 3.1 reddening law of Cardelli, Clayton \& Mathis  (1989).  

The optical and 2MASS near-IR colours for the central star are quite red, consistent with a G/K giant, or an early K-type main sequence star.  Since E$(V-I)$ = 1.28\,E$(B-V)$ = 0.28, we can correct the mean $V$-$I$ colour from TASS to $(V-I)_0$ = 0.97.  This corresponds to a spectral type of K1--2.  Similarly, a reddening corrected $J-H$ colour of 0.55 corresponds to a spectral type of K3--4.  The Tycho data suggest an earlier type around mid-G, but the $B_{T}$-magnitude might be influenced by the underlying hot continuum of the ionizing star.  Based on all available data, we adopt a spectral type of G8--K3 for the cool star hereafter.  The luminosity class is indeterminate until spectra are obtained.

\begin{table}
{\footnotesize	
\begin{center}
\caption{Archival multi-wavelength photometry of the central star of K~1-6.}
\medskip
\label{table:K1-6_CS}
\begin{tabular}{lll}
\hline
Band & Magnitude  & Source\\
\hline	
$F_{\rm UV}$  &   13.97 $\pm$ 0.01   &  GALEX \\	
$N_{\rm UV}$  &   14.62 $\pm$ 0.01    &  GALEX \\				
$B_{T}$  &   13.00 $\pm$  0.29  &  Tycho-2 \\						
$V_{T}$  &   12.12  $\pm$ 0.19    &   Tycho-2    \\
$V$  &   12.16  $\pm$ 0.19   & GSC2.3   \\
$V$  &   12.48 $\pm$ 0.36    & TASS-IV   \\
$R$  &   12.29  $\pm$ 0.22    & NSVS   \\
$R$  &   12.15  $\pm$ 0.08    & UCAC3   \\
$I_{C}$  & 11.23 $\pm$ 0.20&  TASS-IV\\					
$J$      & 10.619 $\pm$ 0.021&  2MASS\\			
$H$      & 10.009 $\pm$ 0.018&  2MASS\\					
$K_{s}$  & 9.809  $\pm$ 0.015&  2MASS\\
\hline
$F_{\rm UV}-N_{\rm UV}$    &    $-0.65$ &   GALEX     \\
$B_{T}-V_{T}$  &    0.88 &   Tycho-2     \\
$V-I_{\rm C}$   &  1.25      &    TASS    \\				
$J-H$    & 0.61 $\pm$ 0.03  &    2MASS    \\		
$H-K_{s}$  & 0.20 $\pm$ 0.03  &    2MASS    \\				
\hline
\end{tabular}
\end{center}
}
\end{table}

We also derived an estimate for the visual magnitude of the hot star visible on GALEX images in the following manner.  Using the GALEX $F_{\rm UV}$ ($\lambda_{\rm eff}$ = 1528\AA) and $N_{\rm UV}$ ($\lambda_{\rm eff}$ = 2270\AA) magnitudes for 24 known PN central stars that show little or no extinction (mean $T_{\rm eff}$ = 100kK), we estimate $(F_{\rm UV}-N_{\rm UV})_0$ = $-0.75$ $\pm$ 0.14 and $(N_{\rm UV}-V)_0$ = $-1.8$ $\pm$ 0.2, after correcting the GALEX magnitudes with a standard extinction law (Cardelli et al. 1989).   We keep the $F_{\rm UV}$ and $N_{\rm UV}$ magnitudes in the AB system.  Our derived values are consistent with the synthetic colours given by Vennes, Kawka \& N\'emeth (2010). 

Using the de-reddened $F_{\rm UV}$ and $N_{\rm UV}$ magnitudes for the central star of K~1-6, we estimate $V_0$ $\simeq$ 15.1 and re-redden to get $V$ $\simeq$ 15.8.  This is $\sim$3.3 magnitudes fainter than the cool star, which explains why there is no clear signature of the hot star in the optical and near-IR colours of the system.  The resulting absolute magnitude assuming $D$ = 1.0 kpc (see Section~\ref{sec:distance}) is $M_V$ $\simeq$ +5.1, consistent with the absolute magnitudes of central stars in other PNe (Frew \& Parker 2010a).  This magnitude places the ionizing star near the top of the white dwarf cooling track, consistent with a low-mass central star in a highly evolved PN.

\subsubsection{Stellar variability}\label{sec:variability}
Interestingly, the cool star is variable, based on data from the Northern Sky Variability Survey (NSVS; Wozniak et al. 2004), where it is designated as NSVS~1225362.  The variability was discovered by Usatov \& Nosulchik (2008), who noted a range of $R$ = 11.97 to 12.74  with P = 21 days.  These authors noted the associated X-ray detection and suggested the star was a likely RS CVn variable.

We have independently determined the period using both the available NSVS and TASS data.   Initially the data from NSVS was folded on an initial estimate of the period from the average separation between apparent peaks in the lightcurve.  New periods around this value were trialed and the amount of scatter around the mean curve were visually estimated. The point at which the scatter was minimised was determined to be the most likely period, 21.3 days. The error in this method was determined to be not less than 0.1 day due to its subjective nature.

A more formal approach was then undertaken.  We combined the TASS $V$ and $I$ data to form a synthetic red magnitude, which was found to be offset from the NSVS $R$-band data by 0.40 mag, in the sense the TASS data are fainter.  After correcting to a common zero point, the TASS and NSVS observations were combined to provide a data set of over 770 points spanning 2749 days.  We used the Peranso\footnote{http://www.peranso.com/} software package using an ANOVA method to determine a period of 21.312 $\pm$ 0.008 days.  The periods estimated from each data set are given in Table~\ref{table:period}.   

\begin{figure}
\begin{center}
\includegraphics[scale=0.48]{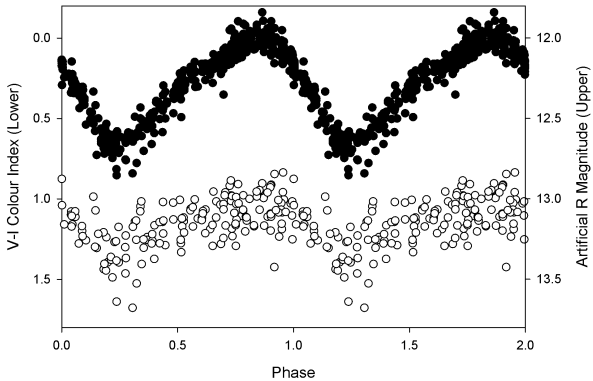}
\caption{Phased light curve of the central star of K~1-6 assuming a period of 21.312 days.  The top graph shows all data points from NSVS (wideband $R$ magnitude) and TASS (synthetic red magnitude --- see text).  The bottom curve shows the TASS $V-I$ colour index as a function of phase.  The star clearly becomes redder as it fades. }
\label{fig:light_curve}
\end{center}
\end{figure}

The combined  light curve as a function of phase is given in Figure~\ref{fig:light_curve}.  The phased-up light-curve is quite regular and is quasi-sinusoidal with an amplitude of $\sim$0.7 mag, but we note a slower rise to maximum followed by a faster decline.   We also looked at the TASS $V-I$ colour index as a function of phase.  The star clearly gets redder as it fades (see the lower plot in Figure~\ref{fig:light_curve}), consistent with either a Cepheid or RS~CVn interpretation.

\begin{table}
{\footnotesize	
\begin{center}
\caption{Estimated period and amplitude of variability}
\medskip
\label{table:period}
\begin{tabular}{lllll}
\hline
Sample & $n$  & Range & Period  & A \\
	& & MJD   & (day) & (mag) \\	
\hline			
NSVS 	 & 604	&  51274 -- 51633  & 21.32 $\pm$ 0.05	& 0.68	\\
TASS 	 & 169 	&  52786 -- 54024  & 21.31 $\pm$ 0.05	 & 0.72	\\
All      & 773	& 51274 -- 54024  & 21.312 $\pm$ 0.008  & 0.70	\\
\hline
\end{tabular}
\end{center}
}
\end{table}

There is no evidence for any eclipses in the available archival data.  We also found no evidence for any rapid variations (on the order of minutes to tens of minutes) from differential photometry of the cool star on our FTN narrow-band and broadband $B$ and $V$ images.  In order to constrain the variability type of the cool star, we considered the following possibilities:\\

\begin{itemize}

\item Irradiation variable -- unlikely as amplitude seems to be too large for period (see De Marco, Hillwig \& Smith 2008);
\item Ellipsoidal variable -- unlikely as amplitude is too large, and light-curve is asymmetric;
\item Small-amplitude red variable (SARV) -- these are red giants of types late K and early M; our amplitude and period are consistent but our star appears too hot.
\item Ordinary pulsating G or K subgiant or giant -- rejected as the observed amplitude is much too large and the period too long;
\item Type II Cepheid  -- appears consistent with observed parameters, but calculated distance derived from the P-L relation is too large;
\item Anomalous Cepheid -- these are metal poor stars and represent the extension of classical Cepheids to lower metallicities and masses -- rejected as our period is too long;
\item Cataclysmic variable -- rejected as there is no evidence of rapid variations on our FTN images, or any eruptive behaviour from archival data;
\item Classical symbiotic star -- rejected as there is no evidence for high-density nebular gas (but we lack spectra) or eruptive behaviour; observed period appears too short.
\item Evolved, chromospherically active variable, e.g. RS CVn or FK Com type -- also feasible but our star has an unusually large amplitude; hard X-ray emission is diagnostic.



\end{itemize}

At first glance the amplitude, period and dereddened $V-I$ colour of the star appear to be consistent with it being a Type II Cepheid.  The light curve has a slow rise and a faster decline which is unusual, but not unprecendented amongst this class (see Figure 3 of Soszy\'nski et al. 2008).  The period falls in the distribution gap between the W~Virginis and RV~Tauri subclasses, but the light curve is more like a W~Vir star, as it lacks the alternating deep and shallow minima characteristic of the RV Tau class (Wallerstein 2002).

However, the detection of hard X-rays (see section~\ref{sec:x-ray}) strongly suggests that it is a coronally active cool star, such as an RS~CVn system, as first suggested by Usatov \& Nosulchik (2008).  It may in fact be a member of the V471~Tau subclass, or even an FK~Com star, pending spectroscopic analysis.   We note however that the observed amplitude ($\sim$0.7 mag) is quite large for either a FK~Com or RS~CVn star.

\subsection{Distance}\label{sec:distance}

If the cool component seen in K~1-6 was a Type~II Cepheid, the distance would be easy to estimate since these stars follow a well defined period--luminosity (PL) relation, which is 1.5--2.0 mags fainter at a given period than the classical Cepheid PL relation (Soszy\'nski et al. 2008).  On the assumption that the star is a W~Vir star, we calculate a distance using the mean reddening-corrected $V$-band magnitude from the TASS and the relation of Pritzl et al. (2003).  We obtain a distance of 6.3 $\pm$ 0.6~kpc.   However, we immediately discount this interpretation because the resulting stellar transverse velocity would be unphysical ($\sim$900\,\kms) and the associated PN would be impossibly massive ($>$10\,$M_{\odot}$).  Furthermore, the $z$-distance from the Galactic plane would be 2.3\,kpc, placing the PN well outside the dust and gas layer in the disk (Spitzer 1978; Dickey \& Lockman 1990), hence a strong ISM interaction would not be expected.   In addition, Type II Cepheids are rare stars and the chance of having one in a physical binary with a PN central star is remote.  Lastly, we note that Type II Cepheids are not strong X-ray sources. 

While we have discounted the cool star as a Type~II Cepheid, we can still estimate an approximate photometric parallax from a combination of the reddening-corrected $V-I$ and $J-H$ colours.  The Tycho-2 data are not reliable at this magnitude and we also decided against using the $K_{s}$-band magnitude from 2MASS, as it may be affected by hot dust --- note that the star appears to have a weak detection in archival 12 $\mu$m IRAS data, but it is not listed as an IRAS source.

A preliminary estimate of the spectral type from the reddening-corrected colours is G8 -- K3 (see above).  However, we cannot derive the luminosity class from the available data.  Since this is uncertain, a range of possible luminosity classes is given in Table~\ref{table:cool}.  We adopt the absolute magnitudes for each class from Schmidt-Kaler (1982) and Cox (2000) in column~2.  Using the reddening-corrected mean $V_0$ = 11.8 $\pm$ 0.2, we derive the spectroscopic distances in column~3. and the height above the Galactic plane in column~4.   We calculate the transverse velocity (in \kms) for each case using 4.74 $\mu$\,$D$, where $D$ is in pc and $\mu$ is in arcsec\,yr$^{-1}$.  These are given in column~5.  We can immediately rule out the cool star being a main sequence star (the velocity is too low to form a pronounced bow shock in the PN) or a bright giant of luminosity class II (transverse velocity is unphysical, as for the Type II Cepheid interpretation).  A luminosity class of  either III or IV is consistent with the available data.

\begin{table}
{\footnotesize	
\begin{center}
\caption{Inferred properties for the cool component of K~1-6 as a function of spectral class.}
\medskip
\label{table:cool}
\begin{tabular}{llccc}
\hline
Spectral type& $M_V$ &	$D$     & $z$   & $v_{t}$  	\\
	   & 	     & (pc)     & (pc)  & (\kms)  	\\				
K3 V	   &   +6.7  &  105	& 38	&   14		\\ 
K1 V	   &   +6.2  &  130	& 48	&   18		\\ 
G8/K0 IV   &   +3.1  &  550	& 200	&   75		\\ 
K0 III-IV  &   +1.9  &  960	& 350	&   130		\\ 
G8 III	   &   +0.8  & 1580	& 580	&   220		\\ 
K0 III	   &   +0.7  & 1660	& 605	&   230		\\ 
K3 III	   &   +0.3  & 2000	& 730	&   275		\\ 
\hline
\end{tabular}
\end{center}
}
\end{table}

We can use an alternative approach to find the distance, based on the properties of the nebula, assuming it to be a bona fide planetary.  Our recently developed \ha\ surface brightness--radius (SB-$r$) relation (Frew \& Parker 2006; Frew 2008) uses the reddening-corrected \ha\ surface brightness to infer the radius of the PN (in pc).  The angular size of the PN then directly gives the distance. Adopting the dimensions, \ha\ flux, and reddening of the PN from Table~\ref{table:basic}, we derive a distance of 1.6\,kpc using the preliminary SB-$r$ relation of Pierce et al. (2004), or 1.1\,kpc using a short trend applicable to lower-mass shells (Frew \& Parker 2006, 2007).  If we factor in representative 1-sigma uncertainties  on the SB-$r$ distances (Frew 2008), the extreme distance range is 0.8--2.0 kpc.  Since the X-ray properties (see Section~\ref{sec:x-ray}) suggest the nearer end of this range is more likely, we adopt a working distance of 1.0\,kpc for K~1-6 hereafter, and the inferred luminosity class for the cool star is then III-IV.

\subsection{X-ray analysis}\label{sec:x-ray}

\begin{figure*}
\begin{center}
\includegraphics[width=16cm]{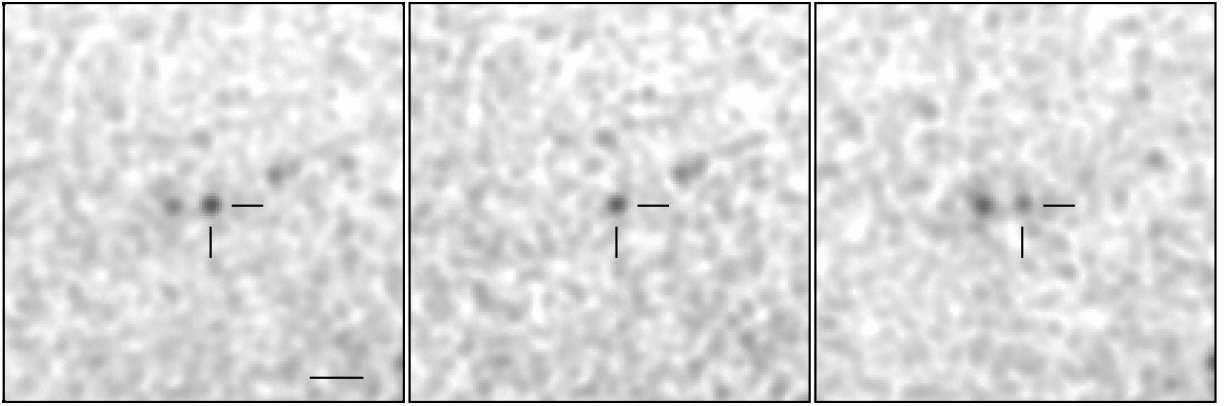}
\caption{Smoothed \emph{ROSAT} PSPC images of the field containing K\,1-6 in the energy ranges 0.1-1.3 keV ({\it left}), 0.1-0.45 keV ({\it center}), and 0.55-1.3 keV ({\it right}). The position of K\,1-6 is marked.
The adjacent source towards the east is likely to be a background AGN, as suggested from its energy hardness.
North is at the top, and east to the left. The scale bar on the first image represents 10\arcmin.}
\label{fig:ROSAT}
\end{center}
\end{figure*}

\begin{figure}
\begin{center}
\includegraphics[width=8.2cm]{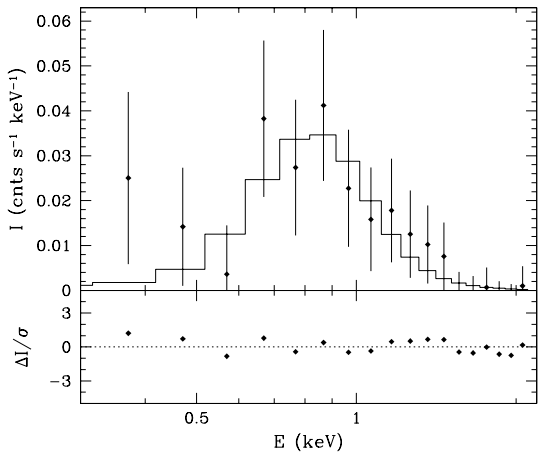}
\caption{ROSAT PSPC spectrum and best-fit model for the central star of K~1-6.  The spectrum can be deconvolved into two components:  a softer component corresponding to the high energy tail of a hot stellar photosphere, and a harder component that is associated with emission from the cool active companion.}\label{fig:Xray_spec}
\end{center}
\end{figure}

The central star of K\,1-6 coincides with the ROSAT source 1RXS\,J200413.0+742533 (Figure~\ref{fig:ROSAT}).  We extracted the ROSAT X-ray spectrum using the HEASOFT software package (Figure~\ref{fig:Xray_spec}).  The background-subtracted X-ray spectrum shows a prominent peak at 0.7--0.9 keV.  Adopting a column density $N_{\rm H}$=1.3$\times$10$^{21}$ cm$^{-2}$ consistent with the assumed $E(B-V)$ = 0.22, the best-fit, thin plasma MEKAL model (Mewe, Kaastra \& Liedahl 1995) has a temperature $kT$ = 0.5$^{+0.4}_{-0.2}$\,keV (or 6$\times$10$^{6}$\,K).  We find that the soft X-ray component is poorly fitted at the adopted extinction, so the asymptotic SFD reddening may be an overestimate.  

The unabsorbed X-ray flux in the 0.3--2.0 keV energy band is 3.1$\times$10$^{-13}$ erg~cm$^{-2}$~s$^{-1}$ with an uncertainty within a factor of two.  At a distance of 1.0\,kpc, it implies an X-ray intrinsic luminosity of $L_{x}$ = 3.7$\times$10$^{31}$($d$/kpc)$^2$ erg~s$^{-1}$.   This luminosity is consistent with an RS~CVn or FK Com star, but near the upper limit for these classes (Walter \& Bowyer 1981; Makarov 2003; Gondoin 2005).   We also note that the luminosity $L_{x}$ is somewhat too bright for the measured $kT$ after referring to the log\,$L_{x}$ versus log\,$T$ plot of Schmitt et al. (1990).  This might also be an indication that the adopted distance is somewhat too large and/or the extinction has been overestimated.

Assuming our X-ray temperature to be correct, we can also use it to place constraints on the nature of the binary star.  The temperature appears too low to be a CV with an accretion disk (Eracleous, Halpern \& Patterson 1991), and even though the combination of low $kT$ and high $L_{x}$ overlaps with symbiotic stars (e.g. M\"urset,  Wolff \& Jordan 1995), there is as yet no corroborating observational evidence that this is indeed a symbiotic system.  The nebular morphology is also unlike the strongly axisymmetric (toroidal and bipolar) forms seen in true symbiotic outflows (Corradi 2003).  However, spectroscopy of the central star might reveal high-density emission lines indicative of the new class of EGB~6-like nuclei (Frew \& Parker 2010).

The absolute bolometric magnitude of the cool component assuming $D$ = 1.0\,kpc is $M_{\rm bol}$ = +1.4 (equivalent to $L_{\rm bol}$ = 22$L_{\odot}$).  The distance-independent ratio, $R_{x}$ = log\,(L$_x$/L$_{\rm bol}$) is $\sim-3.4$ which is close to the saturated level of X-ray emission from an active corona (G\"udel, Guinan \& Skinner 1997; G\"udel 2004; G\"udel \& Naz\'e 2009). 

Our observed photometric modulation has a 21.3 day period (Section~\ref{sec:variability}), but it remains uncertain what this period represents.  Assuming $T_{\rm eff}$ $\sim$5000~K for the cool star, the observed luminosity leads to a radius of $\sim$6.5~$R_{\odot}$.  Furthermore, if we assume a mass of $\sim$1.0--1.5~$M_{\odot}$, the fundamental pulsation period is calculated to be only 0.4--0.6\,days, so pulsation does not appear to be the explanation of the observed light variation.  

Since the cool star appears to be a subgiant or giant, we cannot use a standard rotation-activity relation applicable to main-sequence stars (e.g. Pallavicini et al. 1981; G\"udel et al. 1997) to determine the rotation period from the $L_{x}$ value.  However, the observed period is consistent with the observed rotation periods of other coronally active cool giants.  Gondoin (2005, 2007) has shown there is a relationship beween the X-ray surface flux and rotation for cool giants, both single and binary.  Gondoin (2007) adopts the following relation:

\begin{equation}
\label{eq:gondoin}
{\rm log}(F_{x}) = -0.95\,{\rm log}(P_{\rm rot}) + 7.1
\end{equation}

For the cool star in K~1-6, the X-ray surface flux is $\sim$10$^7$~erg\,s$^{-1}$cm$^{-2}$, good to a factor of two.  The predicted rotation period using equation~\ref{eq:gondoin} is then only $\sim$1 day, which must be in error as this exceeds the critical rotation rate.  Better data, including an independent determination of the star's rotation period, are needed to see if the star follows the relationship between activity and rotation. We note that two active giants with comparable log\,$F_{\rm x}$ values, FF~Aqr and 39~Cet, have rotation periods of 9.2 and 77.2 days respectively, and fall well away from the relation of Gondoin (2007); curiously, both have hot subluminous companions.

In many respects this system resembles the old, high-latitude PN LoTr~5 (Longmore \& Tritton 1980; Graham et al. 2004), which holds a binary nucleus with a hot WD and a chromospherically active, optically variable G5\,III companion, IN~Comae, which is a strong X-ray emitter (Guerrero et al. 2000; Montez et al. 2010).  However, the optical amplitude of the variable cool component in K~1-6 is much larger.

\subsection{PN/ISM interaction}\label{sec:interaction}

We note several PNe with similar stratified, one-sided morphologies to K~1-6.  Examples of such `sosie' PNe are Lo~10 (Schwarz, Corradi \& Melnick 1992), BV~5-2 (Manchado et al. 1996), Te~3 (Jacoby et al. 2010), IsWe~1 (Xilouris et al. 1996;  Tweedy \& Kwitter 1996), and Abell~31 (Abell 1966; Tweedy \& Kwitter 1994).  The morphologies of these nebulae are attributed to an ISM interaction in each case.   We show now that this explanation is valid to explain the morphology seen in K~1-6.

At our preferred distance of 1.0\,kpc, the transverse velocity of the binary central star is 140 \kms, which is at the high-velocity tail of the velocity distribution for the local thin disk population (Skuljan, Hearnshaw \& Cottrell 1999).  The true space motion might be higher (the radial velocity is currently unknown) so it might also be a thick disk object or a rare halo PN.  Either way, the high velocity explains the morphological asymmetry of the PN, due to a strong interaction with the ISM.   Borkowski, Sarazin \& Soker (1990) showed that significant distortion of a PN becomes apparent once the pressure of the expanding PN becomes equal with the ram pressure of the surrounding warm ISM.  Following Borkowski et al. (1990) and Jacoby \& Van de Steene (1995), this relation is expressed mathematically as:

\begin{equation}
\label{eq:pressure}
n_0\,(v_{\ast} + v_{\rm exp})^2  = n_e\,v_{\rm exp}^2
\end{equation}

For our preferred distance of 1.0\,kpc, the $z$-height is 365\,pc.  Using the best-fit distribution of HI with height in the Galaxy (Dickey \& Lockman 1990), the ISM density is $n_0$ = 0.06\,cm\,$^{-3}$ at this $z$-height.  We have no kinematic information on K~1-6, so we assume an expansion velocity of 20 \kms\ for the shell (Weinberger 1989), and the root-mean-square electron density of the PN is calculated as $n_e$ = 20\,cm\,$^{-3}$ based on the observed diameter and flux.  

Re-arranging equation~\ref{eq:pressure} and solving for $v_{\ast}$, we infer a space velocity of 350 \kms\ which is consistent to roughly a factor of two with the estimated transverse velocity at this distance.  At smaller distances the ISM density is higher, and the required space velocity is reduced (but remember the transverse velocity scales with $D$).  Alternatively, the ISM is denser than predicted by Dickey \& Lockman (1990) along this sight-line.  Of course the central star could also have a significant radial velocity of a similar magnitude to the transverse component, in which case the star belongs to the halo.  Considering the assumptions of this rather coarse approach, we conclude that the observed velocity of the star is enough to explain the observed nebular interaction with the ISM at this $z$-height.

\section{Future Work}\label{sec:future}

Initially, medium-resolution, optical spectroscopy of the central binary (or ternary) star of K~1-6 is required in order to remove any ambiguity in its classification, and to determine its metallicity.  Ultimately, an accurate spectral type and luminosity class for the central star will allow an improved photometric parallax for this interesting PN.  If the distance is confirmed, then K~1-6 is an addition to the solar neighbourhood PN census (Frew \& Parker 2010).  This analysis might be hinting that there are other nearby, low-mass PN remaining to be discovered, especially well off the Galactic plane.


A combined time-series photometric and radial velocity study will put constraints on the physical parameters of the central binary system.  If the cool star is an RS CVn variable, we would predict strong Ca\,II H and K emission and Mg\,II emission which are signatures of strong chromospheric activity.  Higher resolution spectroscopic observations would provide direct evidence for the cool star having experienced spin-up.  Fast rotation in an old star could be a signature of mass accretion from a former companion in a detached binary system, now seen as the ionizing source of the PN.  Jeffries \& Stevens (1996) have proposed that a detached secondary in a wide binary can accrete part of the slow massive wind from an AGB companion.  The gain in mass and angular momentum can spin up the secondary to a state of very rapid rotation; as the AGB star evolves to a white dwarf, the companion now appears as a wind-accretion induced rapid rotator (or WIRRing star).  Soker \& Kastner (2002) showed that this is viable on theoretical grounds, for cases where the initial separation was less than 30--60 AU.  

The cool star might also possibly be an FK~Comae star (Bopp \& Stencel 1981) in a wide binary with the ionizing star.  If this is the case, intense Ca\,II H and K and Mg\,II emission is again predicted, as is direct evidence of rapid rotation.  Unlike the RS~CVn stars, no large radial velocity variations would be expected.

An accurate systemic radial velocity will also provide the necessary constraints on the space motion so that K~1-6 can be definitively classified as either a disk or halo PN.  In the future, an accurate parallax will come from the GAIA\footnote{http://www.rssd.esa.int/gaia/} astrometric satellite.  This space mission, with its microarcsecond positions, proper motions and parallaxes, along with its ability to determine accurate radial velocities, promises to revolutionise our understanding of the Milky Way's stellar content, PNe included (see van Leeuwen 2007, for a fuller discussion).

\section{Conclusions}\label{sec:conclusions}

We obtained narrow-band images of the little studied emission nebula, K 1-6 with the FTN. A full search using online tools uncovered a range of multiwavelength archival data  on the nebula and its central star.  The study was conducted in conjunction with Year~11 high school students as part of an Australian Research Council Linkage Grant science education project, denoted Space To Grow.  As Percy et al. (2008) have emphasised, an important goal is to allow students to ``learn science by doing science'' with real data sets.

Our FTN \ha\ (+ \NII) images show K~1-6  to be a faint elliptical nebula with dimensions 197\arcsec\ $\times$ 160\arcsec\ that shows an asymmetric morphology indicative of an interaction with the ISM.  As detailed in Section 4.2, the {\it available body of evidence} (Frew \& Parker 2010) suggests the nebula is best interpreted as an evolved, low-surface-brightness PN based on its limb-brightened bow-shocked morphology, centrally concentrated \OIII\ emission, ionized shell mass, vertical distance from the Galactic plane, and the observed brightness of its ionizing star.

The off-centred ``central'' star has the photometric colours of a G/K subgiant or giant star, while archival GALEX images show a very hot subdwarf or pre-white dwarf coincident in position with this cool star. The central star is unusual, being strongly variable in the optical with a period of 21.312 $\pm$  0.008 days and an amplitude of 0.7\,mag.  It inhabits a position in period--amplitude space that appears to be unique amongst variable PN central stars (cf.  De Marco et al. 2008).

Based on the stellar colours and X-ray properties, we prefer a RS~CVn classification for the cool star, but the amplitude is unusual for the class. The presence of a hot ionizing source in the system implies that the system may be ternary, as most RS CVn systems contain two cool stars. Alternatively, the star is a member of the V471~Tau subclass,  which is a binary containing an evolved cool star and a hot white dwarf.  A system containing a FK~Com star and a wide hot companion is also a possibility.  Assuming a distance of 1.0\,kpc, the inferred visual magnitude of the ionizing star is consistent with known PN central stars (Frew \& Parker 2010) and it appears to be located on the white dwarf cooling track in the HR diagram.  

The high transverse velocity of the binary central star marks this as a kinematically old system.  The cool star must have been spun up to rotate at a sufficient speed to show strong X-ray emission, which is normally a signature of stellar youth.  The cool star has either been spun up by mass transfer in a close binary or via wind accretion in a wider binary system.  Hence, the companion should show signs of enrichment in s-process elements such as barium, that were derived from the AGB wind;  the central star of the evolved PN Abell~70 (Miszalski et al. 2010b) is just such a case.  However, if the cool component is an FK Com star formed by a binary merger, then no enrichment signature will be seen.

We urge further observations of this interesting system, including deep spectroscopic follow-up of the nebula
and in particular, medium resolution spectroscopy of its central binary (or ternary) star in order to remove
any ambiguity in its classification.

\section*{Acknowledgments}

The observations and data analyses reported in this paper were conducted in conjunction with Year~11 high school students as part of the ARC Linkage Grant {\it Space To Grow} science education project jointly conducted by professional astronomers, educational researchers, teachers, and high-school students. This project provided guaranteed observing time with the Faulkes Telescopes, and was supported under Australian Research Council's Linkage Industry Projects funding scheme (project number LP0989264).  M.A.G. is partially funded by the grants AYA2008-01934 of the Spanish Ministerio de Ciencia e Innovaci\'on (MICINN), and PR2009-0342 of the Spanish Ministerio de Educaci\'on (MEC).

We made use of the Vizier service, Aladin Sky Atlas and SIMBAD database from the CDS, Strasbourg.  We acknowledge the wide range of publicly available online databases, such as 2MASS, GALEX, NSVS, TASS, and VTSS, without which this paper could not be written.  We thank Wayne Rosing for financial support at LCOGT, and thank Edward Gomez, Rachel Street, Tim Brown, Tim Lister and Eric Saunders for assistance.  We also thank Simon O'Toole for useful discussions.

\section*{References}

\noindent Abell, G.O., 1966, ApJ, 144, 259\\

\noindent Acker, A., Ochsenbein, F., Stenholm, B., Tylenda, R., Marcout, J., Schohn, C., 1992,  Strasbourg-ESO Catalogue of Galactic Planetary Nebulae (Garching: ESO)\\

\noindent Bopp, B.W., Stencel, R.E., 1981, ApJ, 247, L131\\

\noindent Borkowski, K.J., Sarazin, C.L., Soker, N., 1990, ApJ, 360, 173\\

\noindent Cahn, J.H., Kaler, J.B., 1971, ApJS, 22, 319\\

\noindent Cardelli, J.A., Clayton, G.C., Mathis, J.S., 1989, ApJ, 345, 245\\

\noindent Cohen, J.G., Rosenthal, A.J. 1983, ApJ, 268, 689\\

\noindent Condon, J.J. et al., 1998, AJ, 115, 1693\\

\noindent Cox, A.N., 2000, Allen's Astrophysical Quantities, 4th edition (New York: AIP Press \& Springer)\\

\noindent Corradi, R.L.M., 2003, ASPC, 303, 393\\

\noindent Cutri, R.M. et al., 2003, VizieR On-line Data Catalog: II/246\\

\noindent De Marco, O., 2009, PASP, 121, 316\\

\noindent De Marco, O., Hillwig, T.C., Smith, A.J., 2008, AJ, 136, 323\\

\noindent Dennison, B., Simonetti, J.H., Topasna, G.A., 1998,  PASA, 15, 14\\

\noindent Dickey, J.M., Lockman, F.J., 1990, ARA\&A, 28, 215\\

\noindent Drew, J.E. et al., 2005, MNRAS, 362, 753\\

\noindent Droege, T.F., Richmond, M.W., Sallman, M.P., Creager R.P., 2006, PASP, 118, 1666\\

\noindent Ellis, G.L., Grayson, E.T., Bond, H.E. 1984, PASP, 96, 283\\

\noindent Eracleous, M., Halpern, J., Patterson, J., 1991, ApJ, 382, 290\\

\noindent Finkbeiner, D.P., 2003, ApJS, 146, 407\\

\noindent Frew, D.J., 2008, PhD thesis, Macquarie University\\

\noindent Frew, D.J., Parker, Q.A., 2006, in IAUS 234, Planetary Nebulae in our Galaxy and Beyond, ed. M.J. Barlow, R.H. M\'endez (Cambridge: CUP), p. 49\\

\noindent Frew, D.J., Parker, Q.A., 2007, in APN IV Proceedings, ed. A. Manchado, R.L.M Corradi, N. Soker, IAC Electronic Pub., p. 475\\

\noindent Frew, D.J., Parker Q.A., 2010, PASA, 27, 129\\

\noindent Frew, DJ., Madsen, G.J., Parker, Q.A., 2006. In IAUS 234, Planetary Nebulae in our Galaxy and Beyond, ed. M.J. Barlow, R.H. M\'endez (Cambridge: CUP), p. 395\\

\noindent Frew, D.J., Parker Q.A., Russeil, D., 2006, MNRAS, 372, 1081\\

\noindent Frew, D.J., Madsen, G.J., O'Toole, S.J., Parker Q.A., 2010a, PASA, 27, 203\\

\noindent Frew, D.J., et al., 2010b, in preparation\\

\noindent Gaustad, J.E., McCullough, P.R., Rosing, W., Van Buren, D.J., 2001, PASP, 113, 1326\\

\noindent Gon\c{c}alves, D.R., Corradi, R.L.M., Mampaso, A., 2001, ApJ, 547, 302\\

\noindent Goerigk, W., Mebold, U., Reif, K., Kalberla, P.M.W., Velden, L., 1983, A\&A, 120, 63\\

\noindent Gondoin, P., 2005, A\&A, 444, 531\\

\noindent Gondoin, P., 2007, A\&A, 464, 1101\\

\noindent Graham, M.F., Meaburn, J., L\'opez, J.A., Harman, D.J., Holloway, A.J., 2004, MNRAS, 347, 1370\\

\noindent Guerrero, M.A., Chu, Y.-H., Gruendl, R.A., 2000, ApJS, 129, 295\\

\noindent G\"udel, M., 2004, A\&A Rev., 12, 71\\

\noindent G\"udel, M., Naz\'e, Y., 2009, A\&A Rev., 17, 309\\

\noindent G\"udel, M., Guinan, E.F., Skinner, S.L., 1997, ApJ, 483, 947\\

\noindent Haffner, L.M., Reynolds, R.J., Tufte, S.L., Madsen, G.J., Jaehnig, K.P., \& Percival J.W., 2003, ApJS, 149, 405\\

\noindent Hidas, M.G., Hawkins, E., Walker, Z.,  Brown, T.M.,  Rosing, W.E., 2008, AN, 329, 269\\

\noindent Hobbs, G. et al.,  2009, PASA, 26, 468\\

\noindent H$\o$g, E., et al., 2000,  A\&A, 355, L27\\

\noindent Hollis, J.M., Oliversen, R.J., Wagner, R.M., Feibelman, W.A., 1992, ApJ, 393, 217\\

\noindent Hollow, R. et al. 2008, in ASPC, 400, Ed. M.G. Gibbs, J. Barnes, J. G. Manning, and B. Partridge (San Francisco: Astronomical Society of the Pacific), p. 190\\

\noindent Hua, C.T., Kwok, S., 1999, A\&AS, 138, 275\\

\noindent Jacoby, G.H., 1981, ApJ, 244, 903\\

\noindent Jacoby, G.H., Van de Steene, G., 1995, AJ, 110, 1285\\

\noindent Jacoby, G.H., et al., 2010, PASA, 27, 156\\

\noindent Jeffries, R.D., Stevens, I.R., 1996, MNRAS, 279, 180\\

\noindent Kohoutek, L., 1962, BAC, 13, 120\\

\noindent Kohoutek, L., 2001, A\&A, 378, 843\\

\noindent Krautter, J., Klaas, U., Radons, G., 1987,  A\&A, 181, 373\\

\noindent Kwitter K.B., Jacoby G.H., Lydon, T.J., 1988, AJ, 96, 997\\

\noindent Kwok, S., Purton, C.R., Fitzgerald, P.M., 1978, ApJ, 219, L125\\

\noindent Lewis, F., Street, R., Roche, P., Stroud, V., Russell, D.M., 2010, Advances in Astronomy, 2010, 873059\\

\noindent Longmore A.J., Tritton S.B., 1980, MNRAS, 193, L521\\

\noindent Makarov, V.V., 2003, AJ, 126, 1996\\

\noindent Manchado, A., Guerrero, M.A., Stanghellini, L., Serra-Ricart, M., 1996,  The IAC Morphological Catalog of Northern Galactic Planetary Nebulae.  Instituto de Astrof\'isica de Canarias\\

\noindent Martin, D.C. et al. 2005, ApJ, 619, L1\\

\noindent Mewe, R., Kaastra, J.S., Liedahl, D.A., 1995, Legacy, 6, 16\\

\noindent Miszalski, B., Acker, A., Parker, Q. A., Moffat, A.F.J., 2009, A\&A, 505, 249\\

\noindent Miszalski, B., et al., 2010a, Proceedings of the APN5 conference (Ebrary, in press), eprint: arXiv:1009.2890\\

\noindent Miszalski, B., et al., 2010b, in preparation\\

\noindent Monet, D.G., et al., 2004, AJ, 125, 984\\

\noindent Montez, R., Jr., De Marco, O., Kastner, J.H., Chu, Y.-H., 2010, ApJ, 721, 1820\\

\noindent Morrissey, P. et al., 2007, ApJS, 173, 682\\

\noindent M\"urset, U., Wolff, B., Jordan, S., 1995, A\&A, 319, 201\\

\noindent Oostra, B., 2006, The Physics Teacher, 44, 153\\

\noindent Pallavicini, R., Golub, L., Rosner, R., Vaiana, G.S., Ayres, T., Linsky, J.L. 1981, ApJ, 248, 279\\

\noindent Parker, Q.A. et al., 2005, MNRAS, 362, 689\\


\noindent Percy, J.R., Favaro, E., Glasheen J., Ho B., Sato, H., Turner, D.G., 2008, JAAVSO, 36, 145\\

\noindent Perek L., Kohoutek L., 1967, Catalogue of Galactic Planetary Nebulae (Prague: Academia Publishing House)\\

\noindent Pierce, M.J., Frew, D.J., Parker, Q.A., K\"oppen, J., 2004, PASA, 21, 334\\

\noindent Pritzl, B.J., Smith, H.A., Stetson, P.B., Catelan, M., Sweigart, A.V., Layden, A.C., Rich, R.M., 2003, AJ, 126, 1381\\

\noindent Sabin, L., Zijlstra, A.A., Wareing, C., Corradi, R.L.M., Mampaso, A., Viironen, K., Wright, N.J., Parker, Q.A., 2010, PASA, 27, 166\\

\noindent Schlegel, D.J., Finkbeiner, D.P., Davis, M., 1998, ApJ, 500, 525\\

\noindent Schmidt-Kaler, T., 1982,  In Landolt-B\"ornstein: Numerical Data and Functional Relationships in Science and Technology -- New Series, Group 6, Astronomy and Astrophysics, volume 2, p. 1\\

\noindent Schwarz, H.E., Corradi, R.L.M., Melnick, J., 1992, A\&AS, 96, 23\\

\noindent Schmitt, J.H.M.M., Collura, A., Sciortino, S., Vaiana, G.S., Harnden, F.R., Jr., Rosner, R., 1990, ApJ, 365, 704\\

\noindent Skuljan, J., Hearnshaw, J.B., Cottrell, P.L., 1999, MNRAS, 308, 731\\

\noindent Skrutskie, M.F., et al. 2006, AJ, 131, 1163\\

\noindent Soszy\'nski, I., et al., 2008,  AcA, 58, 293\\

\noindent Tweedy R.W., Kwitter, K.B., 1994, AJ 108, 188\\

\noindent Tweedy R.W., Kwitter, K.B., 1996, ApJS 107, 255\\

\noindent Usatov, M., Nosulchik, A., 2008, OEJV, 88, 1\\

\noindent van Leeuwen, F., 2007, Hipparcos, the New Reduction of the Raw Data, Astrophysics and Space Science Library, vol. 350 (Dordrecht:Springer)\\

\noindent Vennes, S., Kawka, A., N\'emeth, P., 2010, MNRAS (in press), eprint: arXiv:1008.3823\\

\noindent Voges, W. et al., 1999, A\&A, 349, 389\\

\noindent Wallerstein, G., 2002, PASP, 114, 689\\

\noindent Walter, F.M., Bowyer, S., 1981, ApJ, 245, 671\\

\noindent Wareing, C.J., 2010, PASA, 27, 220\\

\noindent Wareing, C.J., Zijlstra, A.A., O'Brien, T.J., 2007, MNRAS, 382, 1233\\

\noindent Watson, C., Henden, A.A., Price, A., 2009, VizieR On-line Data Catalog: B/vsx\\

\noindent Wo\'zniak, P.R. et al., 2004, AJ, 127, 2436\\

\noindent Xilouris, K.M., Papamastorakis, J., Paleologou E., Terzian, Y., 1996, A\&A, 310, 603\\

\noindent Zacharias, N., et al., 2010, AJ, 139, 2184\\

\noindent Zijlstra A.A., Pottasch, S.R., Bignell, C., 1989,  A\&AS, 79, 329\\


\end{document}